\documentclass[aps,prb,superscriptaddress,showpacs,reprint]{revtex4-1}

\usepackage[utf8]{inputenc}
\usepackage{amsmath,amsfonts,amssymb}
\usepackage{pifont}
\usepackage{rotating}
\usepackage{enumerate}
\usepackage{filecontents}
\usepackage{graphicx}    
\usepackage{epstopdf}
\usepackage[table,dvipsnames,svgnames,x11names]{xcolor} 
\usepackage{soul} 
\usepackage{subfigure}
\setstcolor{red}

\usepackage{multirow,bigstrut}
\newsavebox{\measurebox}

\usepackage{hyperref}
\definecolor{dark-red}{rgb}{0.9,0.15,0.15}
\definecolor{dark-blue}{rgb}{0.15,0.15,0.4}
\definecolor{dark2-blue}{rgb}{0.15,0.15,0.8}
\definecolor{medium-blue}{rgb}{0,0,0.5}
\hypersetup{
    colorlinks, linkcolor={black},
    citecolor={dark-blue}, urlcolor={medium-blue}
}
\begin{document}

\title{FeRhCrSi: A new spin semimetal with room temperature spin-valve behavior}

\author{Y. Venkateswara}
\thanks{Authors contributed equally}
\affiliation{Magnetic Materials Laboratory, Department of Physics, Indian Institute of Technology Bombay, Mumbai 400076, India}
\affiliation{Spectroscopic Investigations of Novel Systems Laboratory, Department of Physics, Indian Institute of Technology Kanpur, Kanpur 208016, India}

\author{Jadupati Nag}
\thanks{Authors contributed equally}
\affiliation{Magnetic Materials Laboratory, Department of Physics, Indian Institute of Technology Bombay, Mumbai 400076, India}

\affiliation{Graduate School of Advanced Science and Engineering, Hiroshima University, Higashihiroshima, Hiroshima 739-0046, Japan}

\author{S. Shanmukharao Samatham}
\affiliation{Magnetic Materials Laboratory, Department of Physics, Indian Institute of Technology Bombay, Mumbai 400076, India}

\affiliation{Department of Physics, Chaitanya Bharathi Institute of Technology, Gandipet, Hyderabad 500 075, India}

\author{Akhilesh Kumar Patel}
\affiliation{Magnetic Materials Laboratory, Department of Physics, Indian Institute of Technology Bombay, Mumbai 400076, India}

\author{P. D. Babu}
\affiliation{UGC-DAE Consortium for Scientific Research, Mumbai Centre, BARC Campus, Mumbai 400085, India}

\author{Manoj Raama Varma}
\affiliation{National Institute of Interdisciplinary Sciences and Technology (CSIR), Tiruvananthapuram 695019, India}

\author{Jayita Nayak}
\affiliation{Spectroscopic Investigations of Novel Systems Laboratory, Department of Physics, Indian Institute of Technology Kanpur, Kanpur 208016, India}

\author{K. G. Suresh}
\email{suresh@phy.iitb.ac.in}
\affiliation{Magnetic Materials Laboratory, Department of Physics, Indian Institute of Technology Bombay, Mumbai 400076, India}

\author{Aftab Alam}
\email{aftab@phy.iitb.ac.in}
\affiliation{Department of Physics, Indian Institute of Technology Bombay, Mumbai 400076, India}


\begin{abstract}
Spin semimetals are a recently discovered new class of spintronic materials, which exhibit a band gap in one spin channel while a semimetallic feature in the other and thus allows for tunable 
spin transport. Here, we present experimental verification of spin semimetallic behavior in FeRhCrSi, a quaternary Heusler alloy with saturation moment 2 $\mu_B$ and Curie temperature $>$ 400 K. 
It crystallises in the L2$_1$ structure with 50$\%$ antisite disorder between Fe and Rh. Below 300 K, it shows a weakly temperature dependent electrical resistivity with negative temperature 
coefficient, indicating the normal semimetal or spin semimetal behavior. Anomalous magnetoresistance data reveals dominant contribution from asymmetric part, a clear signature of spin-valve nature, which is retained even at room temperature. \textcolor{black}{The asymmetric part of magneto-resistance shows an unusual increase with increasing temperature.} Hall measurements confirm the anomalous nature of conductivity originating from the intrinsic Berry curvature, with holes being the majority carriers. Ab-initio simulation confirms a unique long-range ferrimagnetic ordering to be the ground state, explaining the origin behind the unexpected low saturation moment. The ferrimagnetic disordered structure confirms the spin semimetallic feature of FeRhCrSi, as observed experimentally.
\end{abstract}


\date{\today}
\maketitle
{\it Introduction:}
Since the discovery of half metals\cite{deGroot-HM-prl-theory}, the search for new/potential spintronic materials\cite{ReviewOnSpintronics-Hirohata-JMMM} has been an active area of research. 
Half metals are special kind of magnetic (ferro, ferri- or antiferro-) materials in which one spin band is metallic while the other is semiconducting or insulating. This enables them to generate highly spin polarized charge carriers for their electrical conduction. Recent progress along these lines has witnessed the proposal for a few interesting classes of materials. Among these, three classes have shown great promises, namely (i) spin gapless semiconductors\cite{PhysRevLett.100.156404} e.g. Mn$_2$CoAl\cite{Ouardi-Mn2CoAl-prl-exp-theory}, CoFeMnSi\cite{Bainsla-CoFeMnSi-prb-exp-theory}, CrVTiAl\cite{Ozdogan-LiMgPdSn-QHA-JAP-theory,CrVTiAl-prb-exp-theory,Stephen-CrVTiAl-APL-exp}, (ii) bipolar magnetic semiconductors e.g. VNbRuAl\cite{VNbRuAl-Nag-prb-exp-theory}, and (iii) spin semimetals e.g. FeRhCrGe\cite{FeRhCrGe-prb-exp-theory,FeRhCrZ-Khandy-JAP-theory}, Mn$_2$LiZ (Z=Al and Ga)\cite{Mn2LiZ-Tariq-MRB-theory}. These materials can also produce highly spin polarized charge carriers i.e., either electrons or holes or both for their electrical conduction. Heusler alloys have been extremely fertile for numerous applications such as spintronics, thermoelectric, topotronics etc. In spintronics, however, majority of work has been focused on alloys composed of $3d$ transition elements. Only a few based on $4d$ transition elements are reported e.g. CoRhMnGe\cite{CoRhMnGe-Deepika-prb-exp-theory}, CoRuMnSi\cite{CoRuMnSi-JMMM-exp-theory}, FeRhCrGe\cite{FeRhCrGe-prb-exp-theory,FeRhCrZ-Khandy-JAP-theory}, Fe$_2$RhSi\cite{PhysRevB.104.094402}, and VNbRuAl\cite{VNbRuAl-Nag-prb-exp-theory}. The main advantage of $4d$ transition elements based alloys is the reduced anti-site disorder due to their quite different sizes and electronegativities of the constituent elements, as compared to those of $3d$ based alloys. The replacement by $5d$ elements also has the advantage of inducing strong spin-orbit coupling. Achieving high magnetic ordering temperature $T_C$ (above room temperature) is, however, challenging. Thus,  Heusler alloys containing $4d$ transition elements with high spin polarization(P) as well as high $T_C$ are prospective candidates for future spintronic research.

It is interesting to note that most of the Heusler alloys follow certain empirical rules such as (i) electronegativity of constituent elements dictating ground state crystal configuration, (ii) integer magnetic moment dictated by the total valence electron count (following Slater-Pauling (SP) rule\cite{ozdougan2013slater}), (iii) superconductivity for most of the 27-valence electron systems etc. However, there exist a few alloys such as Ru$_2$MnGe\cite{Ru2MnGe-Gotoh-PBCM-exp,Ru2MnZ-Kanomata-JJAP-exp}, Ru$_2$CrGe\cite{Ru2CrGe-JPCM-exp} where the magnetic moment fails to obey the SP rule, despite having an ordered structure. Ru$_2$MnGe and Ru$_2$CrGe have 27 and 26 valence electrons and hence are expected to show a net magnetization of 3 and 2 $\mu_B$ respectively, as per the SP rule. Both the alloys are reported to be antiferromagnetic (AFM) due to strong super exchange interaction. AFM coupling exists between Mn-Mn ions in Ru$_2$MnGe (Cr-Cr ions in Ru$_2$CrGe) while Mn-Ru (Cr-Ru) pairs maintain a ferromagnetic (FM) coupling. Such contrasting magnetic properties between experiment and empirical prediction exists in several related compounds, but a convincing reason for such a difference is lacking in the literature.

In this letter, we predict spin semimetallic feature along with room temperature spin-valve behavior of FeRhCrSi using a combined experimental and theoretical study. Band structure 
calculation\cite{Feng-FeRhCrSi-AS-theory,Khandy-FeRhCrSi-JAP-theory} shows a net magnetization of 3 $\mu_B$ for ordered FeRhCrSi, in accordance with the SP rule. However, our experimental data reveal a saturation magnetization of 2 $\mu_B$. Such a discrepancy might arise due to various reasons e.g. (1) presence of super-exchange interactions (2) presence of disorder (3) formation of magnetic domains within a given matrix (4) a combination of two or all of these. This forms yet another impetus which prompted us to undertake a thorough investigation of FeRhCrSi to better understand this discrepancy. A careful inspection of the XRD data confirms FeRhCrSi to crystallize in LiMgPdSn type cubic structure with Cr occupying octahedral sites while Fe and Rh share the tetrahedral sites with 50\% antisite disorder. Our ab-initio simulation reveals that it is not only the anti-site disorder between Fe and Rh but a special long range super-exchange AFM interactions between the Cr-atoms which explains the discrepancy between theoretical (3 $\mu_B$) and experimental moments (2 $\mu_B$). Transport measurement confirms a negative temperature coefficient of resistivity indicating a spin semimetallic behavior. Field dependence of resistivity shows anomalous magneto-resistance (MR) with dominant contribution from its asymmetric part, indicating the spin valve nature. Interestingly, the spin valve behavior is retained even at room temperature. Hall measurements indicate the intrinsic Berry curvature mediated anomalous Hall conductivity (AHC), with holes being the majority carriers. Ab-initio simulation confirms the spin semimetallic behavior in the experimental structure of FeRhCrSi with a unique long range ferrimagnetic ordering amongst the magnetic ions. FeRhCrSi is thus special as it shows several complementary properties (spin semimetal, spin valve and anomalous Hall transport) with unconventional magnetization. \textcolor{black}{A recent article studied the electronic/magnetic properties of a large number of $4d$ and $5d$ based quaternary Heusler alloys, some of which may show similar features as that of FeRhCrSi.}\cite{PhysRevMaterials.6.114407}

\begin{figure}[t]
\centering
\includegraphics[width=\linewidth]{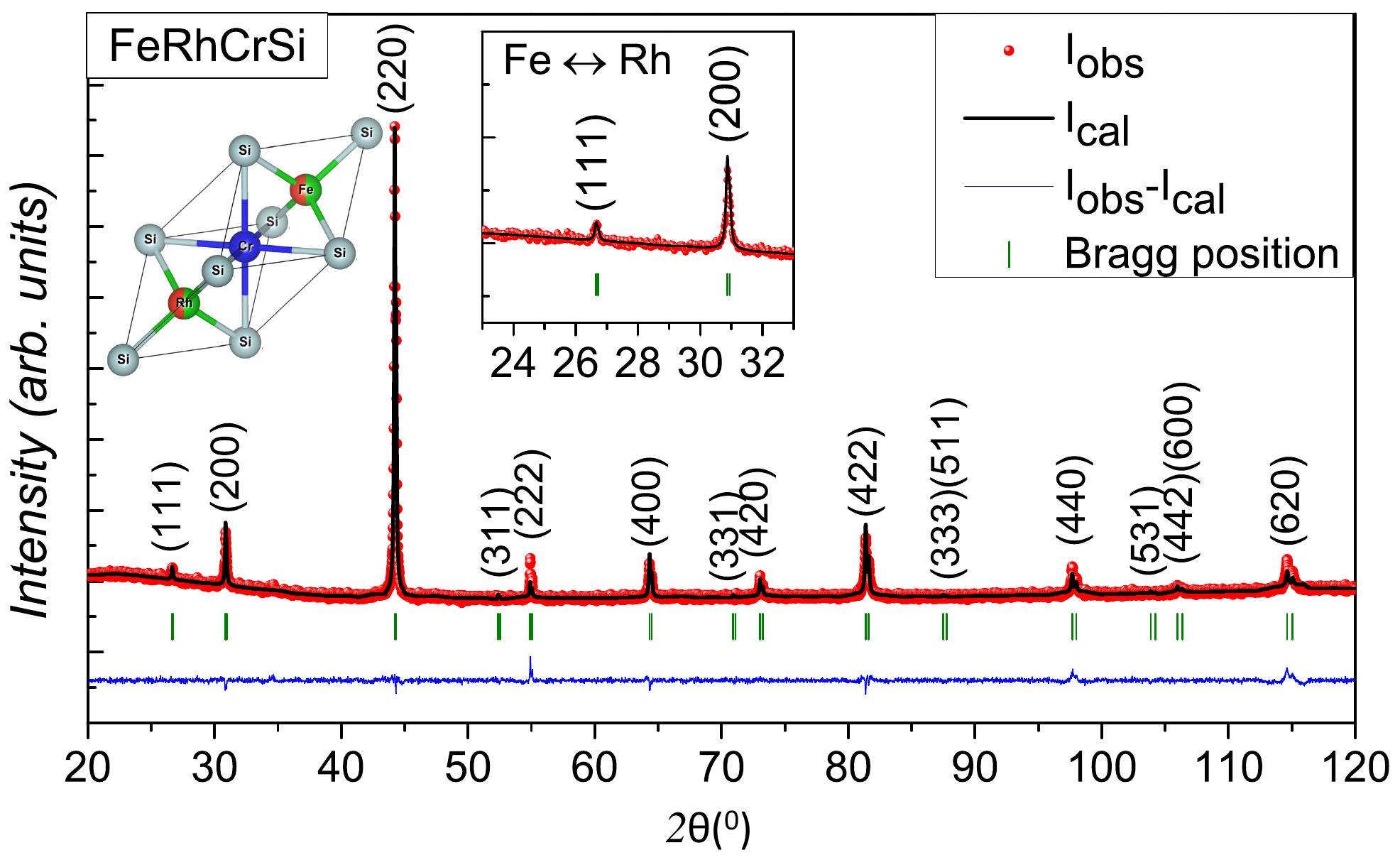}
\caption{ For FeRhCrSi, X-ray diffraction pattern along with the best Rietveld refinement fit with Cr at octahedral site and Fe \& Rh at tetrahedral site with 50$\%$ disorder (corresponding crystal structure in the inset). Another inset shows zoomed-in view of XRD data near (111) \& (200) superlattice reflections.}
\label{fig:XRD-FeRhCrSi}
\end{figure}


{\it Experimental and Computational details:} Polycrystalline FeRhCrSi alloys were prepared by arc-melting method as reported in earlier papers\cite{CrVTiAl-prb-exp-theory}. Further experimental details are given in the supplementary material (SM)\cite{sm}.
First-principles calculations are done using full potential linearised augmented plane-wave (FLAPW) method as implemented in Fleur.\cite{Fl,FleurRef1-PRB,FleurRef2-PRB,FleurRef3-PRB,FleurRef4-PRB} Other computational details are provided in SM\cite{sm}.
FeRhCrSi belongs to a quaternary Heusler alloy (XX$^{'}$YZ) with LiMgPdSn prototype structure (space group $F\bar{4}3m$). The structure can be visualised as 4 interpenetrating fcc sublattices with Wyckoff positions 4a, 4b, 4c and 4d. To find the most stable crystallographic configuration, a 4-atom primitive cell is used. In general, for a quaternary XX$'$YZ alloy, there exists three energetically non-degenerate configurations (keeping Z at 4a site). They are
(1) X at 4d, X$'$ at 4c, and Y at 4b sites  (Type I),
(2) X at 4b, X$'$ at 4c, and Y at 4d sites (Type II),
(3) X at 4c, X$'$ at 4b, and Y at 4d  (Type III).
They are shown in SM\cite{sm}.

{\it Crystal structure:}
Figure \ref{fig:XRD-FeRhCrSi} shows the room temperature X-ray diffraction pattern of FeRhCrSi, indicating a LiMnPdSn type crystal phase with a lattice parameter of 5.79 \AA. Rietveld refinement of XRD pattern with different atomic arrangements are shown in SM\cite{sm}. The best fit was achieved with the Type I configuration in which Cr occupies the octahedral sites  while Fe and Rh share the tetrahedral sites with 50\% disorder. The corresponding primitive cell is shown in the inset of Fig. \ref{fig:XRD-FeRhCrSi}. Another inset shows a zoomed-in view of the refined data near the superlattice reflections (111) and (200). An extremely small intensity mismatch for the reflections (222) and (620) indicates the presence of some texture in the sample. This arises when there are multiple preferred orientations in the structure. Thus, we conclude that FeRhCrSi crystallizes in L2$_1$ structure in Type I configuration with Cr at 4b, Si at 4a and 50\% disorder between Fe and Rh atoms.

{\it Magnetization:} Figure S3(a) of SM\cite{sm} shows the field ($H$) dependence of magnetization ($M$) at $T=$3 K and 300 K.
FeRhCrSi has a saturation magnetization of $\sim$2 $\mu_B/f.u.$ at 3 K. Since it has 27 valence electrons ($N_{val}$), SP rule ($M=N_{val}-24$) predicts 3 $\mu_B$, which clearly indicates 
the presence of antiferromagnetic/ferrimagnetic correlations and/or disorder in the system. Nevertheless, integer moment hints at the possibility of it being a promising spintronic material. Figure S3(b)\cite{sm} shows $M$ vs. $T$ in zero field cooled (ZFC) and field cooled warming (FCW) modes at $H=$ 500 Oe, clearly confirming reasonably high T$_C$.

\begin{figure}[t]
\centering
\includegraphics[width=\linewidth]{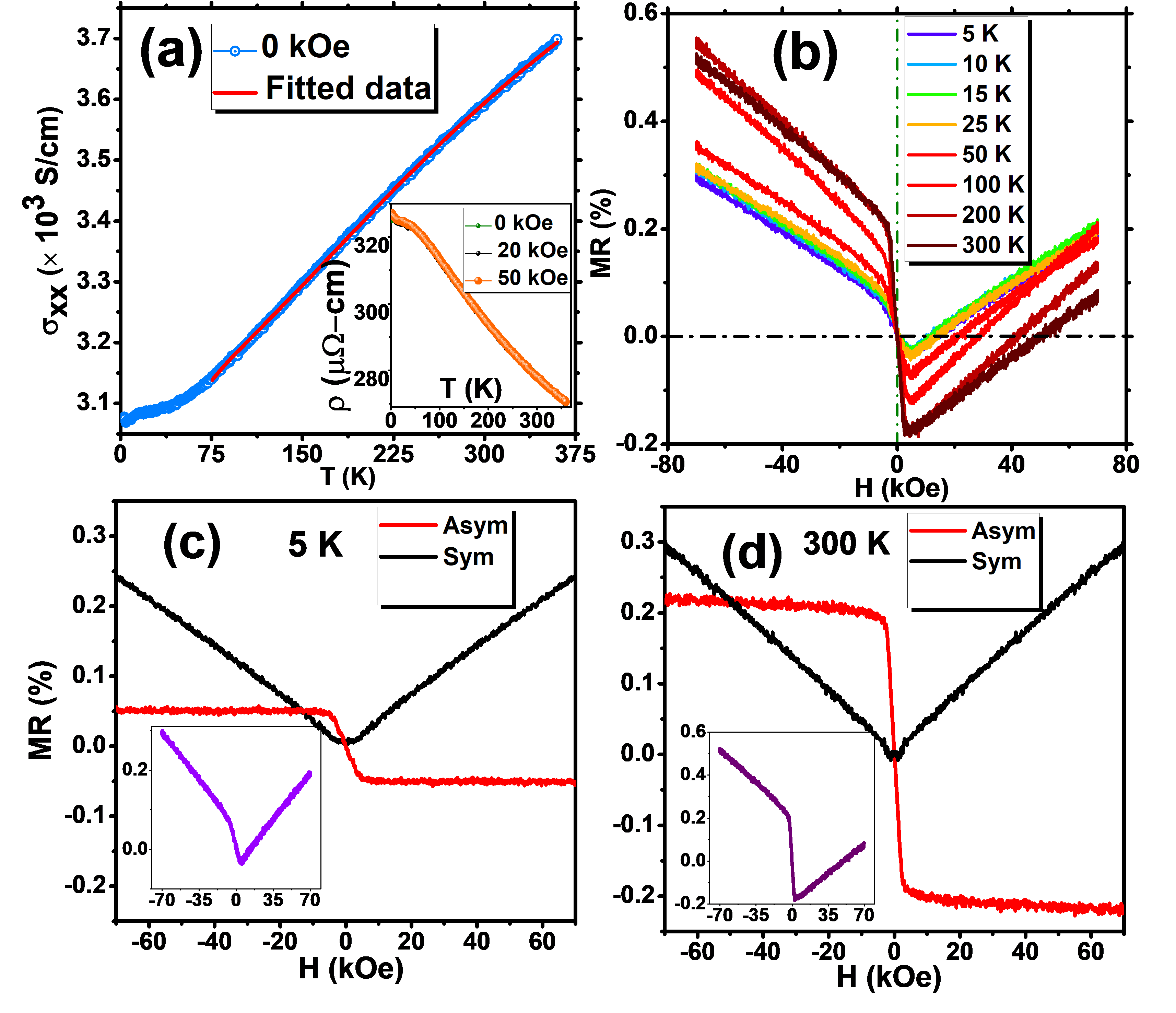}
\caption{For FeRhCrSi (a) longitudinal conductivity ($\sigma_{xx}$) vs. $T$ and two-carrier model fit in zero field. Inset shows the resistivity ($\rho=\rho_{xx}$) vs. $T$ in three different $H$. (b) \textcolor{black}{Total MR (Symmetric+Asymmetric components) vs. $H$ at various $T$. (c)-(d) Asymmetric and symmetric components of MR at 300 and 5 K. Insets show the total MR vs. $H$.}}
\label{fig:R-FeRhCrSi}
\end{figure}

{\it Resistivity: }
Figure \ref{fig:R-FeRhCrSi}(a) shows the conductivity ($\sigma_{xx}$) vs. $T$ at different $H$ for FeRhCrSi. The inset shows resistivity($\rho_{xx}$) vs. $T$ at different $H$. Negative temperature coefficient throughout the $T$ range indicates the semiconducting or semimetallic nature. The order of magnitude and the $T$-variation of $\sigma_{xx}$ are quite comparable to those of semimetallic systems.\cite{FeRhCrGe-prb-exp-theory} Further, a two-carrier model\cite{kittel2007introduction} involving both the charge carriers is used to better understand the transport. The total conductivity can be expressed as, 	$\sigma(T) = e(n_e\mu_e + n_h \mu_h)$, where $n_i=n_{i0} \exp(-\Delta E_i/k_BT) (i=e,h)$ are the carrier concentrations for electrons/holes with mobilities $\mu_i$ and energy gaps $\Delta E_i$. The mobilities can be written as, $\mu_i=(a_iT+b_i)^{-1}=\mu_{i0}/(a'_iT+1)$, where $a$ corresponds to the electron-phonon scattering, while $b$ arises from mobility due to defects at $0$ K. The zero-field $\sigma$ data is fitted with the above equation in the $T$-range 75-350 K giving the energy gaps 37.8 and 0.34 meV, suggesting defect scattering to be the dominant mechanism.

\begin{figure}[t]
\centering
\includegraphics[width=8.6cm,height=7.0cm]{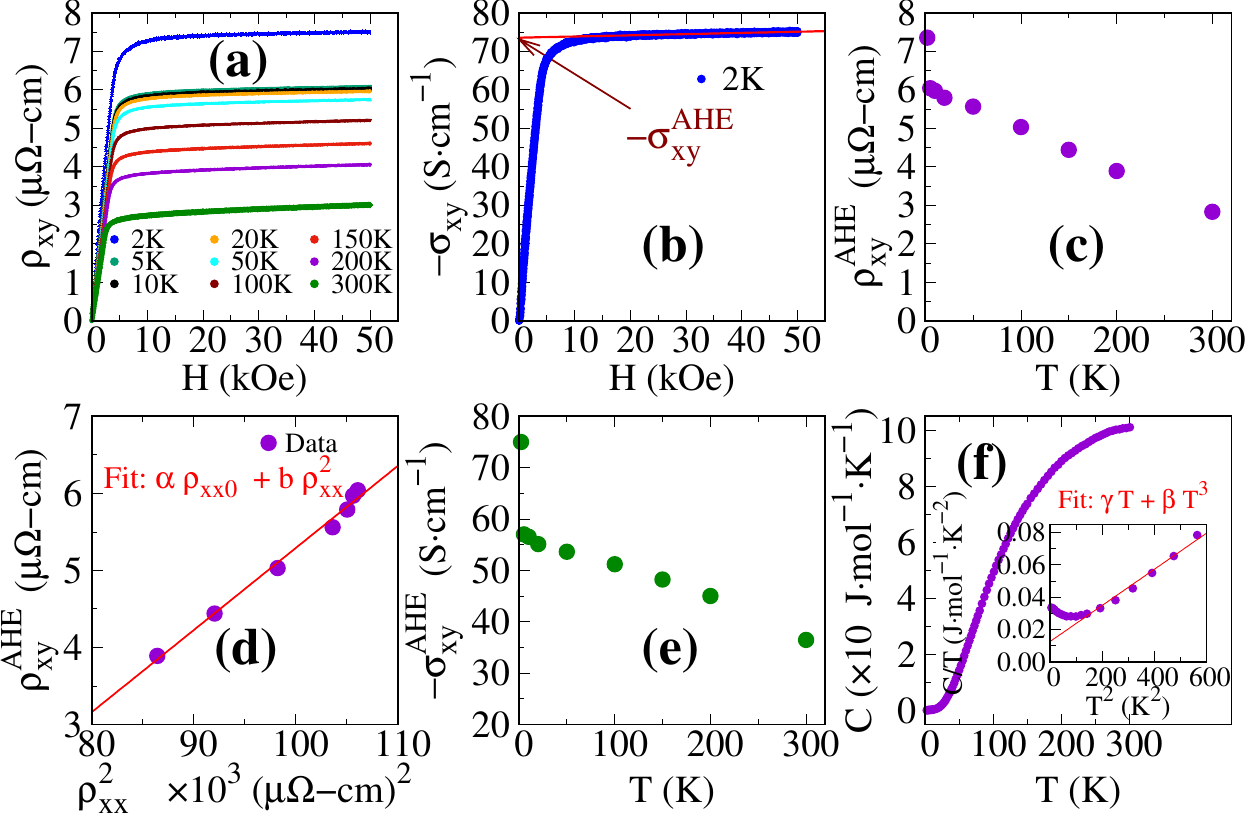}
\caption{For FeRhCrSi (a) Hall resistivity ($\rho_{xy}$) vs. $H$ at various $T$. (b) Hall conductivity ($\sigma_{xy}$) vs. $H$ at 2 K. (c) $\rho_{xy}^{AHE}$ vs. $T$ (d) $\rho_{xy}^{AHE}$ vs. $\rho^2_{xx}$ with linear fit. (e) $\sigma_{xy}^{AHE}$ vs. $T$ (f) Specific heat (C) vs. $T$ in 0 kOe. Inset shows {C/$T$} vs. $T^2$ and linear fit (red line) in low $T$-range.}
\label{fig:H-FeRhCrSi}
\end{figure}

{\it Magnetoresistance: }
Figure \ref{fig:R-FeRhCrSi}(b) shows the field dependence of MR (MR$(H)=\frac{\rho(H)-\rho(0)}{\rho(0)}$) at various $T$. Interestingly, MR shows various anomalies (1) fascinating asymmetric features between positive and negative fields throughout the $T$ range (2) a crossover behavior (negative to positive) with a clear minimum around 4.5 kOe \textcolor{black}{or change of slope at $\sim \pm$4.5 kOe. (3) non-saturating behavior till 70 kOe. The magneto-resistance shows an unconventional behavior with clear rise in magnitude with the increase in temperature. The derivative of MR curves resembles that of M-H curves indicating that the asymmetric part of MR is mainly contributed by the bulk magnetization. To better understand the asymmetric and symmetric parts of the MR, they are calculated using eqs., MR$^{Asym}$=[M(H)-M(-H)]/2 and MR$^{Sym}$=[M(H)+M(-H)]/2 and are shown in Figs. \ref{fig:R-FeRhCrSi}(c) and \ref{fig:R-FeRhCrSi}(d) at 5 K and 300 K respectively. One can notice the dominant asymmetric MR at low $H$. This indicates that the changes within the ferrimagnetic domains at low fields not only dominantly contribute to $M$ but also to its transport properties. As the spins are aligned in one direction with ferrimagnetic ordering within the domains, the charge carriers find lower resistance path for the movement inside the domains as compared to those within domain walls. The region around spin down ions will contribute to pinning effect, similar to spin valve. The symmetric part of MR arises due to nonmagnetic contribution of electrical bands near the Fermi level (E$_F$). The linear rise of symmetric MR with $H$ generally indicates the presence of either electron and hole pockets at E$_F$. The anomalous features indicate unique spin-valve nature\cite{PhysRevB.105.144409,PhysRevLett.109.246601} of FeRhCrSi, with prominent unusual increase in MR with temperature even at room $T$.} These observations suggest ferrimagnetism in the alloy.

{\it \textcolor{black}{Hall conductivity:}}
Figure \ref{fig:H-FeRhCrSi}(a) shows the Hall resistivity ($\rho_{xy}$) vs. $H$ at different $T$. Figure \ref{fig:H-FeRhCrSi}(b) shows the Hall conductivity (calculated using, $\sigma_{xy}$=-$\rho_{xy}/(\rho^2_{xy}$+$\rho^2_{xx}$)\cite{hurd2012hall} vs. H at 2 K. Typically, $\rho_{xy}$ in ferro- or ferrimagnetic materials is expressed 
as  $\rho_{xy}=\rho_{xy}^{AHE}+\rho_{xy}^{OHE}$=$R_{AHE} M+R_{O}H$, where $\rho_{xy}^{AHE}$ and $\rho_{xy}^{OHE}$ are the anomalous and the ordinary Hall resistivities with coefficients $R_{AHE}$ and $R_{O}$ respectively. In these materials, $\rho_{xy}$ is usually dominated by the anomalous Hall effect at low $H$ due to the rapid change in magnetization. Hence, $R^{AHE}$ is obtained by extrapolating the high $H$ data ($>$30 kOe) to zero-field (y-intercept) in the $\rho_{xy}$ curve (Fig.\ref{fig:H-FeRhCrSi}(c)). Similarly, the anomalous Hall conductivity has been calculated and plotted in Fig.\ref{fig:H-FeRhCrSi}(e). To understand the AHE and underlying mechanism (intrinsic and extrinsic), we have used a scaling model reported by Tian \textit{et. al}\cite{PhysRevLett.103.087206}, according to which $\rho^{AHE}$ is expressed as
$\rho^{AHE} = a\rho_{xx0} + a' \rho^2_{xx0} + b\rho^2_{xx}$.
Separating the $T$-independent and dependent terms, this equation takes the form
$\rho^{AHE} = \alpha \rho_{xx0} + b\rho^2_{xx}$,
where $\alpha=a+a'\rho_{xx0}$ is the $T$-independent term representing the extrinsic part arising from the skew scattering and the side jump impurity scattering, while $b$ originates from the intrinsic part. $\rho^{AHE}$ vs. $\rho^2_{xx}$ along with the fit to the last equation is shown in Fig. \ref{fig:H-FeRhCrSi}(d). From the fitting, we obtained $\alpha\sim -0.2\times10^{-2}$ and $b\sim$ 109 S/cm which indicate that $AHE$ is dominated by the intrinsic Berry phase contribution while negative value of $\alpha$ indicates that the extrinsic component is opposite to the Karplus-Luttinger term ($b$).\cite{Karplus-Luttinger-HallEffect-prb} The intrinsic anomalous Hall conductivity is $\sigma_{xy}^{int}=-b\sim-109$ S/cm. The positive slope of high field $\rho_{xy}$ data (above 20 kOe) confirms holes to be the majority charge carriers.

{\it Specific heat:}
Figure \ref{fig:H-FeRhCrSi}(f) shows the specific heat (C) vs. $T$ in zero field. The low-$T$ C-data have been fitted with the equation, C$(T)=$ $\gamma$$T + $$\beta$$T^3$, where first
and second term indicate electronic and low-$T$ phonon contributions respectively. Inset shows {C/$T$} vs. $T^2$ along with linear fit as per the above equation $\gamma$ and $\beta$ obtained from this fit yields the density of states at $E_F$\cite{FeRhCrGe-prb-exp-theory} i.e. $n(E_F)=5.42$ states/eV-f.u. and Debye temperature, $\theta_D$= 412 K. This value of $n(E_F)$ matches fairly well with the simulated value (described in next section) as well as those reported for other semimetals.

\begin{table}[t]
\centering
\caption{For FeRhCrSi, relaxed lattice parameters ($a_{eq}$), atom-projected and total moments, and relative energies of Type I, II and III structural configurations. }
\begin{tabular}{l c c c c c c }
\hline \hline 	
\multirow{3}{*}{Config. } & \multirow{3}{*}{$a_{eq}$(\AA)} & \multicolumn{4}{c}{Moment ($\mu_B$)}  &  \multirow{3}{*}{$\Delta E$(meV/atom)} \\
& & & & & \\
&			 	&\ \	4d		&	4b	&	4c		&\ \ 	Total & 	\\ \hline
&			&\ \	$\mathrm{Fe}$	&$\mathrm{Cr}$	&	$\mathrm{Rh}$ &\  \  	& \\

I  &   5.81 &\ \		0.66			&	2.00			&		0.22		&\  \  3.00	& 	0.0 \\

&			&\ \	$\mathrm{Cr}$	&$\mathrm{Fe}$	&	$\mathrm{Rh}$ & 	& \\
II	 & 5.82	&\  \		-0.31			&	2.46			&		0.31		& \ \ 2.52	& 7.0	\\
&			&\ \	$\mathrm{Cr}$	&$\mathrm{Rh}$	&	$\mathrm{Fe}$ & 	& \\
III	 & 5.81	&\ \		-0.62			&	0.28			&		1.94		& \ \ 1.48	& 217.9	\\
\hline \hline

\end{tabular}
\label{tab:total_energies}
\end{table}

\begin{figure}[t]
\centering
\includegraphics[width=1.0\linewidth]{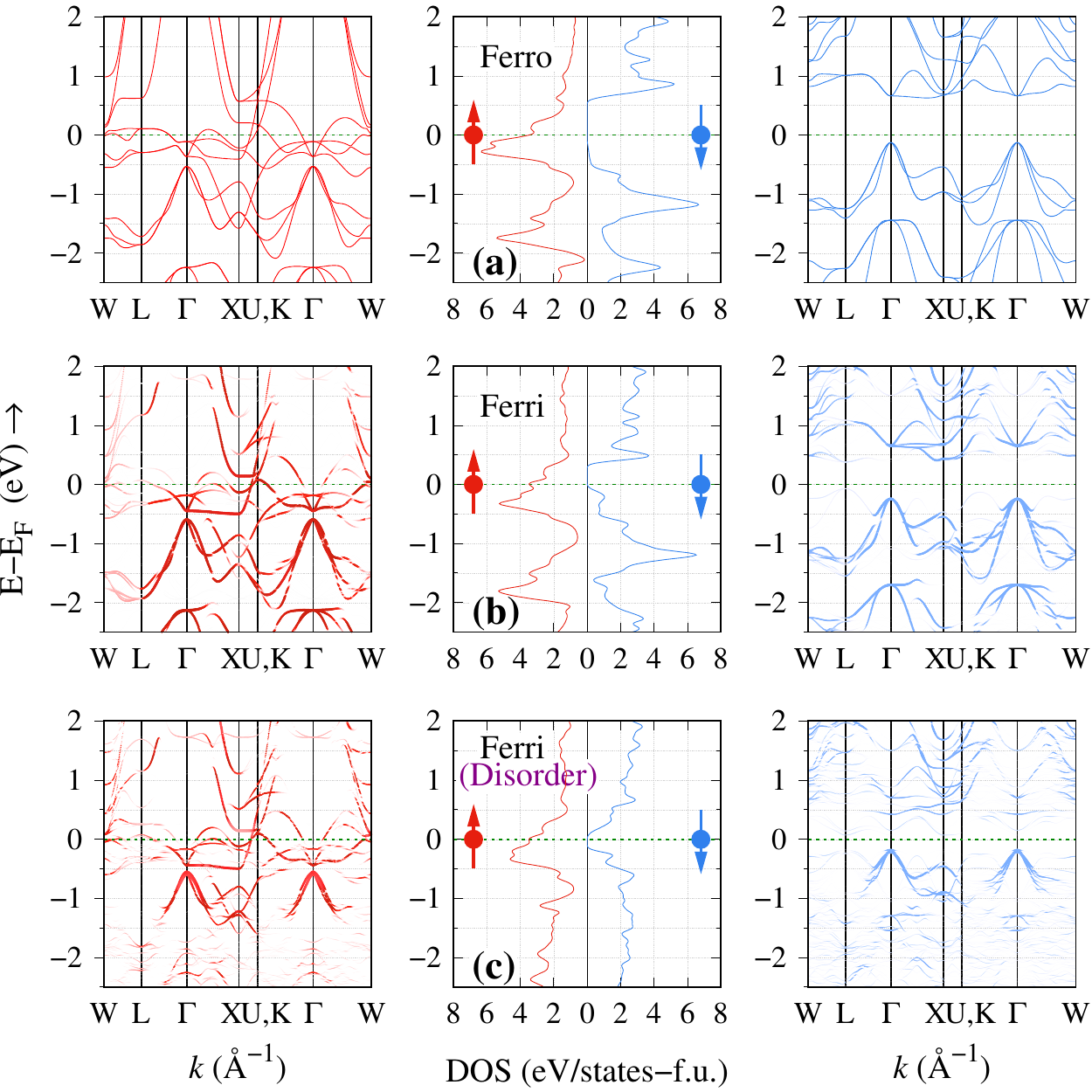}
\caption{Spin resolved band structure and density of states of (a) ferromagnetic (b) ordered ferrimagnetic and (c) disordered ferrimagnetic FeRhCrSi. For comparison, band structures of ordered (16-atom) and disordered (32-atoms SQS) ferrimagnetic (FI) states are unfolded to its primitive cell.}
\label{fig:DB-FeRhCrSi}
\end{figure}

{\it Theoretical results: }
Table \ref{tab:total_energies} shows the relaxed lattice constants, total and atom-projected moments and relative energies of three ordered atomic configurations. Type-I configuration is found to be energetically most favorable with lattice constant 5.81\AA (agreeing fairy well with experiment, 5.79\AA). The constituent elements in this configuration are ferromagnetically aligned with the net magnetization 3.00 $\mu_B$/f.u., which clearly disagrees  with the experimental value of 2.00 $\mu_B$/f.u. This may be due to the long range magnetic interactions and/or the inherent disorder in the system (as predicted from XRD data). To check the former, we considered a 16 atom conventional cell corresponding to Type-I configuration which contains 4-atoms of each type (Fe, Rh, Cr and Ge).  We simulated various ferri- and antiferromagnetic configurations among the magnetic elements. A ferrimagnetic configuration with one of the Cr atoms aligned antiparallel to 3 Cr, 4 Fe and 4 Rh atoms is found to be energetically most favorable. This structure is described by the space group P$\bar{4}$3m (\# 215). More details about this space group is presented in SM\cite{sm}.
This unique magnetic ordering is 2 meV lower in energy as compared to the Type-I FM configuration (shown in Table I), and acquire a net magnetization of 2.00 $\mu_B$/f.u. Spin up Rh, Fe and Cr atoms in this configuration contains a moment of 0.16, 0.88 and 2.02 $\mu_B$ respectively while  spin down Cr atom (henceforth labeled as Cr2) acquire -2.23 $\mu_B$. There is a possibility of clustering or segregation tendency of Cr2 atoms. In order to check this, we have chosen a 32 atom unit cell and simulated two configurations (A) where Cr2 atoms are placed closest to each other and (B) Cr2 atoms placed farthest from each other. These are shown in Fig. (S5,S6) of SM\cite{sm}. It turns out that magnetic configuration (A) is energetically most favorable (by 5 meV), with a net magnetization of 2 $\mu_B$/f.u.

To capture the effect of disorder (as predicted experimentally), we have generated a 32-atom special quasi random structure (SQS) with the same magnetic configuration (A) but site occupancies of Fe and Rh atoms are randomised (see Fig. S5 of SM\cite{sm}). Figure \ref{fig:DB-FeRhCrSi} shows the spin resolved band structure and density of states (DOS) of FM (4-atom primitive cell), ordered and disordered ferrimagnetic (FI) structures. The band structures of the later two cases are unfolded to primitive cell for comparison. All the three structures show spin semimetallic feature with finite DOS(E$_F$) in spin up and finite gap (E$_g$) in spin down channel. The spin down E$_g$ of ordered/disordered FI state is relatively smaller than the FM state due to the flipping of Cr-spins and inherent disorder. The $e_g$ orbitals of Cr2 play a crucial role in reducing the  E$_g$. The predicted spin semimetallic behaviour is consistent with the measured transport data.

{\it \textcolor{black}{Conclusion:} }
We report FeRhCrSi as a new potential candidate to the recently discovered spin semimetal family, in which one spin channel shows semimetallic behavior while the other allows for tunable charge carrier concentrations. Apart from spin semimetallic feature, FeRhCrSi also show promise for room temperature spin valve behavior which is verified by a combined theoretical and experimental study. It crystallizes in LiMgPdSn prototype structure ($F\bar{4}3m$) with L2$_1$ disorder (50-50) between Fe and Rh sites, with lattice constant 5.79\AA. Magnetization measurement confirms a saturation moment of 2.0 $\mu_B$ at 3 K, and the Curie temperature $>$400 K, making it suitable for room temperature applications. Transport measurement reveals negative temperature coefficient of resistivity, indicating the semimetallic or spin semimetallic nature of the alloy. Anomalous magneto-resistance shows dominant contribution from asymmetric part for low fields, indicating the spin valve nature, which survives even at room temperature.\textcolor{black}{The asymmetric part of MR shows an unusual increase in magnitude with increasing $T$.} Hall measurements confirm the presence of anomalous Hall conductivity arising from intrinsic Berry curvature and holes are the majority carriers. Ab-initio simulation reveals a unique long-ranged ferrimagnetic ordering to be the ground state, explaining the origin behind the observed moment of 2$\mu_B$/f.u. (in contrast to the SP rule). The ferrimagnetic, disordered structure confirms the spin semimetallic nature of FeRhCrSi, as observed experimentally. This unique class of materials will have a considerable impact in the field of spintronics as it opens up new  possibilities for physical phenomena and devices based on spin transport.


{\it Acknowledgments:}
YV and JN contributed equally. YV acknowledges Dr. Durgesh Singh for some of the experimental help. YV and JN acknowledge the financial support provided by IIT Bombay. AA acknowledges DST-SERB (Grant No. CRG/2019/002050) for funding to support this research.


\bibliographystyle{apsrev4-2}
\bibliography{bib}

\begin{thebibliography}{34}%
\makeatletter
\providecommand \@ifxundefined [1]{%
 \@ifx{#1\undefined}
}%
\providecommand \@ifnum [1]{%
 \ifnum #1\expandafter \@firstoftwo
 \else \expandafter \@secondoftwo
 \fi
}%
\providecommand \@ifx [1]{%
 \ifx #1\expandafter \@firstoftwo
 \else \expandafter \@secondoftwo
 \fi
}%
\providecommand \natexlab [1]{#1}%
\providecommand \enquote  [1]{``#1''}%
\providecommand \bibnamefont  [1]{#1}%
\providecommand \bibfnamefont [1]{#1}%
\providecommand \citenamefont [1]{#1}%
\providecommand \href@noop [0]{\@secondoftwo}%
\providecommand \href [0]{\begingroup \@sanitize@url \@href}%
\providecommand \@href[1]{\@@startlink{#1}\@@href}%
\providecommand \@@href[1]{\endgroup#1\@@endlink}%
\providecommand \@sanitize@url [0]{\catcode `\\12\catcode `\$12\catcode
  `\&12\catcode `\#12\catcode `\^12\catcode `\_12\catcode `\%12\relax}%
\providecommand \@@startlink[1]{}%
\providecommand \@@endlink[0]{}%
\providecommand \url  [0]{\begingroup\@sanitize@url \@url }%
\providecommand \@url [1]{\endgroup\@href {#1}{\urlprefix }}%
\providecommand \urlprefix  [0]{URL }%
\providecommand \Eprint [0]{\href }%
\providecommand \doibase [0]{https://doi.org/}%
\providecommand \selectlanguage [0]{\@gobble}%
\providecommand \bibinfo  [0]{\@secondoftwo}%
\providecommand \bibfield  [0]{\@secondoftwo}%
\providecommand \translation [1]{[#1]}%
\providecommand \BibitemOpen [0]{}%
\providecommand \bibitemStop [0]{}%
\providecommand \bibitemNoStop [0]{.\EOS\space}%
\providecommand \EOS [0]{\spacefactor3000\relax}%
\providecommand \BibitemShut  [1]{\csname bibitem#1\endcsname}%
\let\auto@bib@innerbib\@empty
\bibitem [{\citenamefont {de~Groot}\ \emph {et~al.}(1983)\citenamefont
  {de~Groot}, \citenamefont {Mueller}, \citenamefont {Engen},\ and\
  \citenamefont {Buschow}}]{deGroot-HM-prl-theory}%
  \BibitemOpen
  \bibfield  {author} {\bibinfo {author} {\bibfnamefont {R.~A.}\ \bibnamefont
  {de~Groot}}, \bibinfo {author} {\bibfnamefont {F.~M.}\ \bibnamefont
  {Mueller}}, \bibinfo {author} {\bibfnamefont {P.~G.~v.}\ \bibnamefont
  {Engen}},\ and\ \bibinfo {author} {\bibfnamefont {K.~H.~J.}\ \bibnamefont
  {Buschow}},\ }\href {https://doi.org/10.1103/PhysRevLett.50.2024} {\bibfield
  {journal} {\bibinfo  {journal} {Phys. Rev. Lett.}\ }\textbf {\bibinfo
  {volume} {50}},\ \bibinfo {pages} {2024} (\bibinfo {year}
  {1983})}\BibitemShut {NoStop}%
\bibitem [{\citenamefont {Hirohata}\ \emph {et~al.}(2020)\citenamefont
  {Hirohata}, \citenamefont {Yamada}, \citenamefont {Nakatani}, \citenamefont
  {Prejbeanu}, \citenamefont {Diény}, \citenamefont {Pirro},\ and\
  \citenamefont {Hillebrands}}]{ReviewOnSpintronics-Hirohata-JMMM}%
  \BibitemOpen
  \bibfield  {author} {\bibinfo {author} {\bibfnamefont {A.}~\bibnamefont
  {Hirohata}}, \bibinfo {author} {\bibfnamefont {K.}~\bibnamefont {Yamada}},
  \bibinfo {author} {\bibfnamefont {Y.}~\bibnamefont {Nakatani}}, \bibinfo
  {author} {\bibfnamefont {I.-L.}\ \bibnamefont {Prejbeanu}}, \bibinfo {author}
  {\bibfnamefont {B.}~\bibnamefont {Diény}}, \bibinfo {author} {\bibfnamefont
  {P.}~\bibnamefont {Pirro}},\ and\ \bibinfo {author} {\bibfnamefont
  {B.}~\bibnamefont {Hillebrands}},\ }\href
  {https://doi.org/https://doi.org/10.1016/j.jmmm.2020.166711} {\bibfield
  {journal} {\bibinfo  {journal} {Journal of Magnetism and Magnetic Materials}\
  }\textbf {\bibinfo {volume} {509}},\ \bibinfo {pages} {166711} (\bibinfo
  {year} {2020})}\BibitemShut {NoStop}%
\bibitem [{\citenamefont {Wang}(2008)}]{PhysRevLett.100.156404}%
  \BibitemOpen
  \bibfield  {author} {\bibinfo {author} {\bibfnamefont {X.~L.}\ \bibnamefont
  {Wang}},\ }\href {https://doi.org/10.1103/PhysRevLett.100.156404} {\bibfield
  {journal} {\bibinfo  {journal} {Phys. Rev. Lett.}\ }\textbf {\bibinfo
  {volume} {100}},\ \bibinfo {pages} {156404} (\bibinfo {year}
  {2008})}\BibitemShut {NoStop}%
\bibitem [{\citenamefont {Ouardi}\ \emph {et~al.}(2013)\citenamefont {Ouardi},
  \citenamefont {Fecher}, \citenamefont {Felser},\ and\ \citenamefont
  {K\"ubler}}]{Ouardi-Mn2CoAl-prl-exp-theory}%
  \BibitemOpen
  \bibfield  {author} {\bibinfo {author} {\bibfnamefont {S.}~\bibnamefont
  {Ouardi}}, \bibinfo {author} {\bibfnamefont {G.~H.}\ \bibnamefont {Fecher}},
  \bibinfo {author} {\bibfnamefont {C.}~\bibnamefont {Felser}},\ and\ \bibinfo
  {author} {\bibfnamefont {J.}~\bibnamefont {K\"ubler}},\ }\href
  {https://doi.org/10.1103/PhysRevLett.110.100401} {\bibfield  {journal}
  {\bibinfo  {journal} {Phys. Rev. Lett.}\ }\textbf {\bibinfo {volume} {110}},\
  \bibinfo {pages} {100401} (\bibinfo {year} {2013})}\BibitemShut {NoStop}%
\bibitem [{\citenamefont {Bainsla}\ \emph {et~al.}(2015)\citenamefont
  {Bainsla}, \citenamefont {Mallick}, \citenamefont {Raja}, \citenamefont
  {Nigam}, \citenamefont {Varaprasad}, \citenamefont {Takahashi}, \citenamefont
  {Alam}, \citenamefont {Suresh},\ and\ \citenamefont
  {Hono}}]{Bainsla-CoFeMnSi-prb-exp-theory}%
  \BibitemOpen
  \bibfield  {author} {\bibinfo {author} {\bibfnamefont {L.}~\bibnamefont
  {Bainsla}}, \bibinfo {author} {\bibfnamefont {A.~I.}\ \bibnamefont
  {Mallick}}, \bibinfo {author} {\bibfnamefont {M.~M.}\ \bibnamefont {Raja}},
  \bibinfo {author} {\bibfnamefont {A.~K.}\ \bibnamefont {Nigam}}, \bibinfo
  {author} {\bibfnamefont {B.~S. D. C.~S.}\ \bibnamefont {Varaprasad}},
  \bibinfo {author} {\bibfnamefont {Y.~K.}\ \bibnamefont {Takahashi}}, \bibinfo
  {author} {\bibfnamefont {A.}~\bibnamefont {Alam}}, \bibinfo {author}
  {\bibfnamefont {K.~G.}\ \bibnamefont {Suresh}},\ and\ \bibinfo {author}
  {\bibfnamefont {K.}~\bibnamefont {Hono}},\ }\href
  {https://doi.org/10.1103/PhysRevB.91.104408} {\bibfield  {journal} {\bibinfo
  {journal} {Phys. Rev. B}\ }\textbf {\bibinfo {volume} {91}},\ \bibinfo
  {pages} {104408} (\bibinfo {year} {2015})}\BibitemShut {NoStop}%
\bibitem [{\citenamefont {Özdoğan}\ \emph {et~al.}(2013)\citenamefont
  {Özdoğan}, \citenamefont {Şaşıoğlu},\ and\ \citenamefont
  {Galanakis}}]{Ozdogan-LiMgPdSn-QHA-JAP-theory}%
  \BibitemOpen
  \bibfield  {author} {\bibinfo {author} {\bibfnamefont {K.}~\bibnamefont
  {Özdoğan}}, \bibinfo {author} {\bibfnamefont {E.}~\bibnamefont
  {Şaşıoğlu}},\ and\ \bibinfo {author} {\bibfnamefont {I.}~\bibnamefont
  {Galanakis}},\ }\href {https://doi.org/10.1063/1.4805063} {\bibfield
  {journal} {\bibinfo  {journal} {Journal of Applied Physics}\ }\textbf
  {\bibinfo {volume} {113}},\ \bibinfo {pages} {193903} (\bibinfo {year}
  {2013})}\BibitemShut {NoStop}%
\bibitem [{\citenamefont {Venkateswara}\ \emph {et~al.}(2018)\citenamefont
  {Venkateswara}, \citenamefont {Gupta}, \citenamefont {Samatham},
  \citenamefont {Varma}, \citenamefont {Enamullah}, \citenamefont {Suresh},\
  and\ \citenamefont {Alam}}]{CrVTiAl-prb-exp-theory}%
  \BibitemOpen
  \bibfield  {author} {\bibinfo {author} {\bibfnamefont {Y.}~\bibnamefont
  {Venkateswara}}, \bibinfo {author} {\bibfnamefont {S.}~\bibnamefont {Gupta}},
  \bibinfo {author} {\bibfnamefont {S.~S.}\ \bibnamefont {Samatham}}, \bibinfo
  {author} {\bibfnamefont {M.~R.}\ \bibnamefont {Varma}}, \bibinfo {author}
  {\bibnamefont {Enamullah}}, \bibinfo {author} {\bibfnamefont {K.~G.}\
  \bibnamefont {Suresh}},\ and\ \bibinfo {author} {\bibfnamefont
  {A.}~\bibnamefont {Alam}},\ }\href
  {https://doi.org/10.1103/PhysRevB.97.054407} {\bibfield  {journal} {\bibinfo
  {journal} {Phys. Rev. B}\ }\textbf {\bibinfo {volume} {97}},\ \bibinfo
  {pages} {054407} (\bibinfo {year} {2018})}\BibitemShut {NoStop}%
\bibitem [{\citenamefont {Stephen}\ \emph {et~al.}(2016)\citenamefont
  {Stephen}, \citenamefont {McDonald}, \citenamefont {Lejeune}, \citenamefont
  {Lewis},\ and\ \citenamefont {Heiman}}]{Stephen-CrVTiAl-APL-exp}%
  \BibitemOpen
  \bibfield  {author} {\bibinfo {author} {\bibfnamefont {G.~M.}\ \bibnamefont
  {Stephen}}, \bibinfo {author} {\bibfnamefont {I.}~\bibnamefont {McDonald}},
  \bibinfo {author} {\bibfnamefont {B.}~\bibnamefont {Lejeune}}, \bibinfo
  {author} {\bibfnamefont {L.~H.}\ \bibnamefont {Lewis}},\ and\ \bibinfo
  {author} {\bibfnamefont {D.}~\bibnamefont {Heiman}},\ }\href
  {https://doi.org/10.1063/1.4971826} {\bibfield  {journal} {\bibinfo
  {journal} {Applied Physics Letters}\ }\textbf {\bibinfo {volume} {109}},\
  \bibinfo {pages} {242401} (\bibinfo {year} {2016})}\BibitemShut {NoStop}%
\bibitem [{\citenamefont {Nag}\ \emph {et~al.}(2021)\citenamefont {Nag},
  \citenamefont {Rani}, \citenamefont {Kangsabanik}, \citenamefont {Singh},
  \citenamefont {Venkatesh}, \citenamefont {Babu}, \citenamefont {Suresh},\
  and\ \citenamefont {Alam}}]{VNbRuAl-Nag-prb-exp-theory}%
  \BibitemOpen
  \bibfield  {author} {\bibinfo {author} {\bibfnamefont {J.}~\bibnamefont
  {Nag}}, \bibinfo {author} {\bibfnamefont {D.}~\bibnamefont {Rani}}, \bibinfo
  {author} {\bibfnamefont {J.}~\bibnamefont {Kangsabanik}}, \bibinfo {author}
  {\bibfnamefont {D.}~\bibnamefont {Singh}}, \bibinfo {author} {\bibfnamefont
  {R.}~\bibnamefont {Venkatesh}}, \bibinfo {author} {\bibfnamefont {P.~D.}\
  \bibnamefont {Babu}}, \bibinfo {author} {\bibfnamefont {K.~G.}\ \bibnamefont
  {Suresh}},\ and\ \bibinfo {author} {\bibfnamefont {A.}~\bibnamefont {Alam}},\
  }\href {https://doi.org/10.1103/PhysRevB.104.134406} {\bibfield  {journal}
  {\bibinfo  {journal} {Phys. Rev. B}\ }\textbf {\bibinfo {volume} {104}},\
  \bibinfo {pages} {134406} (\bibinfo {year} {2021})}\BibitemShut {NoStop}%
\bibitem [{\citenamefont {Venkateswara}\ \emph {et~al.}(2019)\citenamefont
  {Venkateswara}, \citenamefont {Samatham}, \citenamefont {Babu}, \citenamefont
  {Suresh},\ and\ \citenamefont {Alam}}]{FeRhCrGe-prb-exp-theory}%
  \BibitemOpen
  \bibfield  {author} {\bibinfo {author} {\bibfnamefont {Y.}~\bibnamefont
  {Venkateswara}}, \bibinfo {author} {\bibfnamefont {S.~S.}\ \bibnamefont
  {Samatham}}, \bibinfo {author} {\bibfnamefont {P.~D.}\ \bibnamefont {Babu}},
  \bibinfo {author} {\bibfnamefont {K.~G.}\ \bibnamefont {Suresh}},\ and\
  \bibinfo {author} {\bibfnamefont {A.}~\bibnamefont {Alam}},\ }\href
  {https://doi.org/10.1103/PhysRevB.100.180404} {\bibfield  {journal} {\bibinfo
   {journal} {Phys. Rev. B}\ }\textbf {\bibinfo {volume} {100}},\ \bibinfo
  {pages} {180404} (\bibinfo {year} {2019})}\BibitemShut {NoStop}%
\bibitem [{\citenamefont {Khandy}\ and\ \citenamefont
  {Chai}(2020{\natexlab{a}})}]{FeRhCrZ-Khandy-JAP-theory}%
  \BibitemOpen
  \bibfield  {author} {\bibinfo {author} {\bibfnamefont {S.~A.}\ \bibnamefont
  {Khandy}}\ and\ \bibinfo {author} {\bibfnamefont {J.-D.}\ \bibnamefont
  {Chai}},\ }\href {https://doi.org/10.1063/1.5139072} {\bibfield  {journal}
  {\bibinfo  {journal} {Journal of Applied Physics}\ }\textbf {\bibinfo
  {volume} {127}},\ \bibinfo {pages} {165102} (\bibinfo {year}
  {2020}{\natexlab{a}})}\BibitemShut {NoStop}%
\bibitem [{\citenamefont {Hadji}\ \emph {et~al.}(2021)\citenamefont {Hadji},
  \citenamefont {Khalfoun}, \citenamefont {Rached},\ and\ \citenamefont
  {Azzouz-Rached}}]{Mn2LiZ-Tariq-MRB-theory}%
  \BibitemOpen
  \bibfield  {author} {\bibinfo {author} {\bibfnamefont {T.}~\bibnamefont
  {Hadji}}, \bibinfo {author} {\bibfnamefont {H.}~\bibnamefont {Khalfoun}},
  \bibinfo {author} {\bibfnamefont {H.}~\bibnamefont {Rached}},\ and\ \bibinfo
  {author} {\bibfnamefont {A.}~\bibnamefont {Azzouz-Rached}},\ }\href
  {https://doi.org/https://doi.org/10.1016/j.materresbull.2021.111461}
  {\bibfield  {journal} {\bibinfo  {journal} {Materials Research Bulletin}\
  }\textbf {\bibinfo {volume} {143}},\ \bibinfo {pages} {111461} (\bibinfo
  {year} {2021})}\BibitemShut {NoStop}%
\bibitem [{\citenamefont {Rani}\ \emph {et~al.}(2017)\citenamefont {Rani},
  \citenamefont {Enamullah}, \citenamefont {Suresh}, \citenamefont {Yadav},
  \citenamefont {Jha}, \citenamefont {Bhattacharyya}, \citenamefont {Varma},\
  and\ \citenamefont {Alam}}]{CoRhMnGe-Deepika-prb-exp-theory}%
  \BibitemOpen
  \bibfield  {author} {\bibinfo {author} {\bibfnamefont {D.}~\bibnamefont
  {Rani}}, \bibinfo {author} {\bibnamefont {Enamullah}}, \bibinfo {author}
  {\bibfnamefont {K.~G.}\ \bibnamefont {Suresh}}, \bibinfo {author}
  {\bibfnamefont {A.~K.}\ \bibnamefont {Yadav}}, \bibinfo {author}
  {\bibfnamefont {S.~N.}\ \bibnamefont {Jha}}, \bibinfo {author} {\bibfnamefont
  {D.}~\bibnamefont {Bhattacharyya}}, \bibinfo {author} {\bibfnamefont {M.~R.}\
  \bibnamefont {Varma}},\ and\ \bibinfo {author} {\bibfnamefont
  {A.}~\bibnamefont {Alam}},\ }\href
  {https://doi.org/10.1103/PhysRevB.96.184404} {\bibfield  {journal} {\bibinfo
  {journal} {Phys. Rev. B}\ }\textbf {\bibinfo {volume} {96}},\ \bibinfo
  {pages} {184404} (\bibinfo {year} {2017})}\BibitemShut {NoStop}%
\bibitem [{\citenamefont {Venkateswara}\ \emph {et~al.}(2020)\citenamefont
  {Venkateswara}, \citenamefont {Rani}, \citenamefont {Suresh},\ and\
  \citenamefont {Alam}}]{CoRuMnSi-JMMM-exp-theory}%
  \BibitemOpen
  \bibfield  {author} {\bibinfo {author} {\bibfnamefont {Y.}~\bibnamefont
  {Venkateswara}}, \bibinfo {author} {\bibfnamefont {D.}~\bibnamefont {Rani}},
  \bibinfo {author} {\bibfnamefont {K.}~\bibnamefont {Suresh}},\ and\ \bibinfo
  {author} {\bibfnamefont {A.}~\bibnamefont {Alam}},\ }\href
  {https://doi.org/https://doi.org/10.1016/j.jmmm.2020.166536} {\bibfield
  {journal} {\bibinfo  {journal} {Journal of Magnetism and Magnetic Materials}\
  }\textbf {\bibinfo {volume} {502}},\ \bibinfo {pages} {166536} (\bibinfo
  {year} {2020})}\BibitemShut {NoStop}%
\bibitem [{\citenamefont {Venkateswara}\ \emph {et~al.}(2021)\citenamefont
  {Venkateswara}, \citenamefont {Samatham}, \citenamefont {Patel},
  \citenamefont {Babu}, \citenamefont {Varma}, \citenamefont {Suresh},\ and\
  \citenamefont {Alam}}]{PhysRevB.104.094402}%
  \BibitemOpen
  \bibfield  {author} {\bibinfo {author} {\bibfnamefont {Y.}~\bibnamefont
  {Venkateswara}}, \bibinfo {author} {\bibfnamefont {S.~S.}\ \bibnamefont
  {Samatham}}, \bibinfo {author} {\bibfnamefont {A.~K.}\ \bibnamefont {Patel}},
  \bibinfo {author} {\bibfnamefont {P.~D.}\ \bibnamefont {Babu}}, \bibinfo
  {author} {\bibfnamefont {M.~R.}\ \bibnamefont {Varma}}, \bibinfo {author}
  {\bibfnamefont {K.~G.}\ \bibnamefont {Suresh}},\ and\ \bibinfo {author}
  {\bibfnamefont {A.}~\bibnamefont {Alam}},\ }\href
  {https://doi.org/10.1103/PhysRevB.104.094402} {\bibfield  {journal} {\bibinfo
   {journal} {Phys. Rev. B}\ }\textbf {\bibinfo {volume} {104}},\ \bibinfo
  {pages} {094402} (\bibinfo {year} {2021})}\BibitemShut {NoStop}%
\bibitem [{\citenamefont {{\"O}zdo{\u{g}}an}\ \emph {et~al.}(2013)\citenamefont
  {{\"O}zdo{\u{g}}an}, \citenamefont {{\c{S}}a{\c{s}}{\i}o{\u{g}}lu},\ and\
  \citenamefont {Galanakis}}]{ozdougan2013slater}%
  \BibitemOpen
  \bibfield  {author} {\bibinfo {author} {\bibfnamefont {K.}~\bibnamefont
  {{\"O}zdo{\u{g}}an}}, \bibinfo {author} {\bibfnamefont {E.}~\bibnamefont
  {{\c{S}}a{\c{s}}{\i}o{\u{g}}lu}},\ and\ \bibinfo {author} {\bibfnamefont
  {I.}~\bibnamefont {Galanakis}},\ }\href@noop {} {\bibfield  {journal}
  {\bibinfo  {journal} {J. App. Phys.}\ }\textbf {\bibinfo {volume} {113}},\
  \bibinfo {pages} {193903} (\bibinfo {year} {2013})}\BibitemShut {NoStop}%
\bibitem [{\citenamefont {Gotoh}\ \emph {et~al.}(1995)\citenamefont {Gotoh},
  \citenamefont {Ohashi}, \citenamefont {Kanomata},\ and\ \citenamefont
  {Yamaguchi}}]{Ru2MnGe-Gotoh-PBCM-exp}%
  \BibitemOpen
  \bibfield  {author} {\bibinfo {author} {\bibfnamefont {M.}~\bibnamefont
  {Gotoh}}, \bibinfo {author} {\bibfnamefont {M.}~\bibnamefont {Ohashi}},
  \bibinfo {author} {\bibfnamefont {T.}~\bibnamefont {Kanomata}},\ and\
  \bibinfo {author} {\bibfnamefont {Y.}~\bibnamefont {Yamaguchi}},\ }\href
  {https://doi.org/https://doi.org/10.1016/0921-4526(95)00138-Y} {\bibfield
  {journal} {\bibinfo  {journal} {Physica B: Condensed Matter}\ }\textbf
  {\bibinfo {volume} {213-214}},\ \bibinfo {pages} {306} (\bibinfo {year}
  {1995})}\BibitemShut {NoStop}%
\bibitem [{\citenamefont {Kanomata}\ \emph {et~al.}(1993)\citenamefont
  {Kanomata}, \citenamefont {Kikuchi}, \citenamefont {Yamauci},\ and\
  \citenamefont {Kaneko}}]{Ru2MnZ-Kanomata-JJAP-exp}%
  \BibitemOpen
  \bibfield  {author} {\bibinfo {author} {\bibfnamefont {T.}~\bibnamefont
  {Kanomata}}, \bibinfo {author} {\bibfnamefont {M.}~\bibnamefont {Kikuchi}},
  \bibinfo {author} {\bibfnamefont {H.}~\bibnamefont {Yamauci}},\ and\ \bibinfo
  {author} {\bibfnamefont {T.}~\bibnamefont {Kaneko}},\ }\href
  {https://doi.org/10.7567/jjaps.32s3.292} {\bibfield  {journal} {\bibinfo
  {journal} {Japanese Journal of Applied Physics}\ }\textbf {\bibinfo {volume}
  {32}},\ \bibinfo {pages} {292} (\bibinfo {year} {1993})}\BibitemShut
  {NoStop}%
\bibitem [{\citenamefont {Brown}\ \emph {et~al.}(2008)\citenamefont {Brown},
  \citenamefont {Gandy}, \citenamefont {Kanomata}, \citenamefont {Kusakari},
  \citenamefont {Sheikh}, \citenamefont {Neumann}, \citenamefont {Ouladdiaf},\
  and\ \citenamefont {Ziebeck}}]{Ru2CrGe-JPCM-exp}%
  \BibitemOpen
  \bibfield  {author} {\bibinfo {author} {\bibfnamefont {P.~J.}\ \bibnamefont
  {Brown}}, \bibinfo {author} {\bibfnamefont {A.~P.}\ \bibnamefont {Gandy}},
  \bibinfo {author} {\bibfnamefont {T.}~\bibnamefont {Kanomata}}, \bibinfo
  {author} {\bibfnamefont {Y.}~\bibnamefont {Kusakari}}, \bibinfo {author}
  {\bibfnamefont {A.}~\bibnamefont {Sheikh}}, \bibinfo {author} {\bibfnamefont
  {K.-U.}\ \bibnamefont {Neumann}}, \bibinfo {author} {\bibfnamefont
  {B.}~\bibnamefont {Ouladdiaf}},\ and\ \bibinfo {author} {\bibfnamefont
  {K.~R.~A.}\ \bibnamefont {Ziebeck}},\ }\href
  {https://doi.org/10.1088/0953-8984/20/45/455201} {\bibfield  {journal}
  {\bibinfo  {journal} {Journal of Physics: Condensed Matter}\ }\textbf
  {\bibinfo {volume} {20}},\ \bibinfo {pages} {455201} (\bibinfo {year}
  {2008})}\BibitemShut {NoStop}%
\bibitem [{\citenamefont {Feng}\ \emph {et~al.}(2018)\citenamefont {Feng},
  \citenamefont {Ma}, \citenamefont {Yang}, \citenamefont {Lin},\ and\
  \citenamefont {Wang}}]{Feng-FeRhCrSi-AS-theory}%
  \BibitemOpen
  \bibfield  {author} {\bibinfo {author} {\bibfnamefont {L.}~\bibnamefont
  {Feng}}, \bibinfo {author} {\bibfnamefont {J.}~\bibnamefont {Ma}}, \bibinfo
  {author} {\bibfnamefont {Y.}~\bibnamefont {Yang}}, \bibinfo {author}
  {\bibfnamefont {T.}~\bibnamefont {Lin}},\ and\ \bibinfo {author}
  {\bibfnamefont {L.}~\bibnamefont {Wang}},\ }\bibfield  {journal} {\bibinfo
  {journal} {Applied Sciences}\ }\textbf {\bibinfo {volume} {8}},\ \href
  {https://doi.org/10.3390/app8122370} {10.3390/app8122370} (\bibinfo {year}
  {2018})\BibitemShut {NoStop}%
\bibitem [{\citenamefont {Khandy}\ and\ \citenamefont
  {Chai}(2020{\natexlab{b}})}]{Khandy-FeRhCrSi-JAP-theory}%
  \BibitemOpen
  \bibfield  {author} {\bibinfo {author} {\bibfnamefont {S.~A.}\ \bibnamefont
  {Khandy}}\ and\ \bibinfo {author} {\bibfnamefont {J.-D.}\ \bibnamefont
  {Chai}},\ }\href {https://doi.org/10.1063/1.5139072} {\bibfield  {journal}
  {\bibinfo  {journal} {Journal of Applied Physics}\ }\textbf {\bibinfo
  {volume} {127}},\ \bibinfo {pages} {165102} (\bibinfo {year}
  {2020}{\natexlab{b}})}\BibitemShut {NoStop}%
\bibitem [{\citenamefont {Nepal}\ \emph {et~al.}(2022)\citenamefont {Nepal},
  \citenamefont {Dhakal}, \citenamefont {Galanakis}, \citenamefont {Winter},
  \citenamefont {Adhikari},\ and\ \citenamefont
  {Kaphle}}]{PhysRevMaterials.6.114407}%
  \BibitemOpen
  \bibfield  {author} {\bibinfo {author} {\bibfnamefont {S.}~\bibnamefont
  {Nepal}}, \bibinfo {author} {\bibfnamefont {R.}~\bibnamefont {Dhakal}},
  \bibinfo {author} {\bibfnamefont {I.}~\bibnamefont {Galanakis}}, \bibinfo
  {author} {\bibfnamefont {S.~M.}\ \bibnamefont {Winter}}, \bibinfo {author}
  {\bibfnamefont {R.~P.}\ \bibnamefont {Adhikari}},\ and\ \bibinfo {author}
  {\bibfnamefont {G.~C.}\ \bibnamefont {Kaphle}},\ }\href
  {https://doi.org/10.1103/PhysRevMaterials.6.114407} {\bibfield  {journal}
  {\bibinfo  {journal} {Phys. Rev. Mater.}\ }\textbf {\bibinfo {volume} {6}},\
  \bibinfo {pages} {114407} (\bibinfo {year} {2022})}\BibitemShut {NoStop}%
\bibitem [{sm(2022)}]{sm}%
  \BibitemOpen
  \href@noop {} {\bibfield  {journal} {\bibinfo  {journal} {See the
  Supplementary Information}\ } (\bibinfo {year} {2022})}\BibitemShut {NoStop}%
\bibitem [{Fl()}]{Fl}%
  \BibitemOpen
  \href@noop {} {\bibinfo  {journal} {See https://www.flapw.de for more
  information about the Fleur code}\ }\BibitemShut {NoStop}%
\bibitem [{\citenamefont {Kurz}\ \emph
  {et~al.}(2004{\natexlab{a}})\citenamefont {Kurz}, \citenamefont {F\"orster},
  \citenamefont {Nordstr\"om}, \citenamefont {Bihlmayer},\ and\ \citenamefont
  {Bl\"ugel}}]{FleurRef1-PRB}%
  \BibitemOpen
\bibfield  {journal} {  }\bibfield  {author} {\bibinfo {author} {\bibfnamefont
  {P.}~\bibnamefont {Kurz}}, \bibinfo {author} {\bibfnamefont {F.}~\bibnamefont
  {F\"orster}}, \bibinfo {author} {\bibfnamefont {L.}~\bibnamefont
  {Nordstr\"om}}, \bibinfo {author} {\bibfnamefont {G.}~\bibnamefont
  {Bihlmayer}},\ and\ \bibinfo {author} {\bibfnamefont {S.}~\bibnamefont
  {Bl\"ugel}},\ }\href {https://doi.org/10.1103/PhysRevB.69.024415} {\bibfield
  {journal} {\bibinfo  {journal} {Phys. Rev. B}\ }\textbf {\bibinfo {volume}
  {69}},\ \bibinfo {pages} {024415} (\bibinfo {year}
  {2004}{\natexlab{a}})}\BibitemShut {NoStop}%
\bibitem [{\citenamefont {Freimuth}\ \emph {et~al.}(2008)\citenamefont
  {Freimuth}, \citenamefont {Mokrousov}, \citenamefont {Wortmann},
  \citenamefont {Heinze},\ and\ \citenamefont {Bl\"ugel}}]{FleurRef2-PRB}%
  \BibitemOpen
  \bibfield  {author} {\bibinfo {author} {\bibfnamefont {F.}~\bibnamefont
  {Freimuth}}, \bibinfo {author} {\bibfnamefont {Y.}~\bibnamefont {Mokrousov}},
  \bibinfo {author} {\bibfnamefont {D.}~\bibnamefont {Wortmann}}, \bibinfo
  {author} {\bibfnamefont {S.}~\bibnamefont {Heinze}},\ and\ \bibinfo {author}
  {\bibfnamefont {S.}~\bibnamefont {Bl\"ugel}},\ }\href
  {https://doi.org/10.1103/PhysRevB.78.035120} {\bibfield  {journal} {\bibinfo
  {journal} {Phys. Rev. B}\ }\textbf {\bibinfo {volume} {78}},\ \bibinfo
  {pages} {035120} (\bibinfo {year} {2008})}\BibitemShut {NoStop}%
\bibitem [{\citenamefont {Kurz}\ \emph
  {et~al.}(2004{\natexlab{b}})\citenamefont {Kurz}, \citenamefont {F\"orster},
  \citenamefont {Nordstr\"om}, \citenamefont {Bihlmayer},\ and\ \citenamefont
  {Bl\"ugel}}]{FleurRef3-PRB}%
  \BibitemOpen
  \bibfield  {author} {\bibinfo {author} {\bibfnamefont {P.}~\bibnamefont
  {Kurz}}, \bibinfo {author} {\bibfnamefont {F.}~\bibnamefont {F\"orster}},
  \bibinfo {author} {\bibfnamefont {L.}~\bibnamefont {Nordstr\"om}}, \bibinfo
  {author} {\bibfnamefont {G.}~\bibnamefont {Bihlmayer}},\ and\ \bibinfo
  {author} {\bibfnamefont {S.}~\bibnamefont {Bl\"ugel}},\ }\href
  {https://doi.org/10.1103/PhysRevB.69.024415} {\bibfield  {journal} {\bibinfo
  {journal} {Phys. Rev. B}\ }\textbf {\bibinfo {volume} {69}},\ \bibinfo
  {pages} {024415} (\bibinfo {year} {2004}{\natexlab{b}})}\BibitemShut
  {NoStop}%
\bibitem [{\citenamefont {Weinert}\ \emph {et~al.}(1982)\citenamefont
  {Weinert}, \citenamefont {Wimmer},\ and\ \citenamefont
  {Freeman}}]{FleurRef4-PRB}%
  \BibitemOpen
  \bibfield  {author} {\bibinfo {author} {\bibfnamefont {M.}~\bibnamefont
  {Weinert}}, \bibinfo {author} {\bibfnamefont {E.}~\bibnamefont {Wimmer}},\
  and\ \bibinfo {author} {\bibfnamefont {A.~J.}\ \bibnamefont {Freeman}},\
  }\href {https://doi.org/10.1103/PhysRevB.26.4571} {\bibfield  {journal}
  {\bibinfo  {journal} {Phys. Rev. B}\ }\textbf {\bibinfo {volume} {26}},\
  \bibinfo {pages} {4571} (\bibinfo {year} {1982})}\BibitemShut {NoStop}%
\bibitem [{\citenamefont {KITTEL}(2007)}]{kittel2007introduction}%
  \BibitemOpen
  \bibfield  {author} {\bibinfo {author} {\bibfnamefont {C.}~\bibnamefont
  {KITTEL}},\ }\href@noop {} {\bibinfo {title} {Introduction to solid state
  physics. 7th editio}} (\bibinfo {year} {2007})\BibitemShut {NoStop}%
\bibitem [{\citenamefont {Nag}\ \emph {et~al.}(2022)\citenamefont {Nag},
  \citenamefont {Rani}, \citenamefont {Singh}, \citenamefont {Venkatesh},
  \citenamefont {Sahni}, \citenamefont {Yadav}, \citenamefont {Jha},
  \citenamefont {Bhattacharyya}, \citenamefont {Babu}, \citenamefont {Suresh},\
  and\ \citenamefont {Alam}}]{PhysRevB.105.144409}%
  \BibitemOpen
  \bibfield  {author} {\bibinfo {author} {\bibfnamefont {J.}~\bibnamefont
  {Nag}}, \bibinfo {author} {\bibfnamefont {D.}~\bibnamefont {Rani}}, \bibinfo
  {author} {\bibfnamefont {D.}~\bibnamefont {Singh}}, \bibinfo {author}
  {\bibfnamefont {R.}~\bibnamefont {Venkatesh}}, \bibinfo {author}
  {\bibfnamefont {B.}~\bibnamefont {Sahni}}, \bibinfo {author} {\bibfnamefont
  {A.~K.}\ \bibnamefont {Yadav}}, \bibinfo {author} {\bibfnamefont {S.~N.}\
  \bibnamefont {Jha}}, \bibinfo {author} {\bibfnamefont {D.}~\bibnamefont
  {Bhattacharyya}}, \bibinfo {author} {\bibfnamefont {P.~D.}\ \bibnamefont
  {Babu}}, \bibinfo {author} {\bibfnamefont {K.~G.}\ \bibnamefont {Suresh}},\
  and\ \bibinfo {author} {\bibfnamefont {A.}~\bibnamefont {Alam}},\ }\href
  {https://doi.org/10.1103/PhysRevB.105.144409} {\bibfield  {journal} {\bibinfo
   {journal} {Phys. Rev. B}\ }\textbf {\bibinfo {volume} {105}},\ \bibinfo
  {pages} {144409} (\bibinfo {year} {2022})}\BibitemShut {NoStop}%
\bibitem [{\citenamefont {Singh}\ \emph {et~al.}(2012)\citenamefont {Singh},
  \citenamefont {Rawat}, \citenamefont {Muthu}, \citenamefont {D'Souza},
  \citenamefont {Suard}, \citenamefont {Senyshyn}, \citenamefont {Banik},
  \citenamefont {Rajput}, \citenamefont {Bhardwaj}, \citenamefont {Awasthi},
  \citenamefont {Ranjan}, \citenamefont {Arumugam}, \citenamefont {Schlagel},
  \citenamefont {Lograsso}, \citenamefont {Chakrabarti},\ and\ \citenamefont
  {Barman}}]{PhysRevLett.109.246601}%
  \BibitemOpen
  \bibfield  {author} {\bibinfo {author} {\bibfnamefont {S.}~\bibnamefont
  {Singh}}, \bibinfo {author} {\bibfnamefont {R.}~\bibnamefont {Rawat}},
  \bibinfo {author} {\bibfnamefont {S.~E.}\ \bibnamefont {Muthu}}, \bibinfo
  {author} {\bibfnamefont {S.~W.}\ \bibnamefont {D'Souza}}, \bibinfo {author}
  {\bibfnamefont {E.}~\bibnamefont {Suard}}, \bibinfo {author} {\bibfnamefont
  {A.}~\bibnamefont {Senyshyn}}, \bibinfo {author} {\bibfnamefont
  {S.}~\bibnamefont {Banik}}, \bibinfo {author} {\bibfnamefont
  {P.}~\bibnamefont {Rajput}}, \bibinfo {author} {\bibfnamefont
  {S.}~\bibnamefont {Bhardwaj}}, \bibinfo {author} {\bibfnamefont {A.~M.}\
  \bibnamefont {Awasthi}}, \bibinfo {author} {\bibfnamefont {R.}~\bibnamefont
  {Ranjan}}, \bibinfo {author} {\bibfnamefont {S.}~\bibnamefont {Arumugam}},
  \bibinfo {author} {\bibfnamefont {D.~L.}\ \bibnamefont {Schlagel}}, \bibinfo
  {author} {\bibfnamefont {T.~A.}\ \bibnamefont {Lograsso}}, \bibinfo {author}
  {\bibfnamefont {A.}~\bibnamefont {Chakrabarti}},\ and\ \bibinfo {author}
  {\bibfnamefont {S.~R.}\ \bibnamefont {Barman}},\ }\href
  {https://doi.org/10.1103/PhysRevLett.109.246601} {\bibfield  {journal}
  {\bibinfo  {journal} {Phys. Rev. Lett.}\ }\textbf {\bibinfo {volume} {109}},\
  \bibinfo {pages} {246601} (\bibinfo {year} {2012})}\BibitemShut {NoStop}%
\bibitem [{\citenamefont {Hurd}(2012)}]{hurd2012hall}%
  \BibitemOpen
  \bibfield  {author} {\bibinfo {author} {\bibfnamefont {C.}~\bibnamefont
  {Hurd}},\ }\href@noop {} {\emph {\bibinfo {title} {The Hall effect in metals
  and alloys}}}\ (\bibinfo  {publisher} {Springer Science \& Business Media},\
  \bibinfo {year} {2012})\ p.~\bibinfo {pages} {42}\BibitemShut {NoStop}%
\bibitem [{\citenamefont {Tian}\ \emph {et~al.}(2009)\citenamefont {Tian},
  \citenamefont {Ye},\ and\ \citenamefont {Jin}}]{PhysRevLett.103.087206}%
  \BibitemOpen
  \bibfield  {author} {\bibinfo {author} {\bibfnamefont {Y.}~\bibnamefont
  {Tian}}, \bibinfo {author} {\bibfnamefont {L.}~\bibnamefont {Ye}},\ and\
  \bibinfo {author} {\bibfnamefont {X.}~\bibnamefont {Jin}},\ }\href
  {https://doi.org/10.1103/PhysRevLett.103.087206} {\bibfield  {journal}
  {\bibinfo  {journal} {Phys. Rev. Lett.}\ }\textbf {\bibinfo {volume} {103}},\
  \bibinfo {pages} {087206} (\bibinfo {year} {2009})}\BibitemShut {NoStop}%
\bibitem [{\citenamefont {Karplus}\ and\ \citenamefont
  {Luttinger}(1954)}]{Karplus-Luttinger-HallEffect-prb}%
  \BibitemOpen
  \bibfield  {author} {\bibinfo {author} {\bibfnamefont {R.}~\bibnamefont
  {Karplus}}\ and\ \bibinfo {author} {\bibfnamefont {J.~M.}\ \bibnamefont
  {Luttinger}},\ }\href {https://doi.org/10.1103/PhysRev.95.1154} {\bibfield
  {journal} {\bibinfo  {journal} {Phys. Rev.}\ }\textbf {\bibinfo {volume}
  {95}},\ \bibinfo {pages} {1154} (\bibinfo {year} {1954})}\BibitemShut
  {NoStop}%
\end{thebibliography}%

\end{document}